\documentclass[
    ,final            
  ]
  {aipproc}

\layoutstyle{8x11double}

\def\lesssim{\mathbin{\lower 3pt\hbox 
      {$\rlap{\raise 5pt\hbox{$\char'074$}}\mathchar"7218$}}} 
\def\gtrsim{\mathbin{\lower 3pt\hbox
      {$\rlap{\raise 5pt\hbox{$\char'076$}}\mathchar"7218$}}}  

\newcommand\sun{\odot}

\begin{document}

\title{The Magnetic Fields of Anomalous X-ray Pulsars}



\author{Feryal \"Ozel}{
  address={Department of Physics, University of Arizona, 1118. E. 4th St.
Tucson, AZ 85704}
}

\author{Tolga G\"uver}{
  address={Istanbul University, Science Faculty, Astronomy \& Space
Sciences Department, Beyaz\i t, Istanbul, 34119}
  ,altaddress={Department of Physics, University of Arizona, 1118. E. 4th St.
Tucson, AZ 85704} 
}

\author{Ersin G\"o\u{g}\"u\c{s}}{
  address={Sabanc\i~University, Faculty of Engineering Natural Sciences, 
34956 Turkey}
}

\begin{abstract}
Anomalous X-ray Pulsars (AXPs) belong to a class of neutron stars
believed to harbor the strongest magnetic fields in the universe, as
indicated by their energetic bursts and their rapid spindowns. We have
developed a theoretical model that takes into account processes in the
atmospheres and magnetospheres of ultramagnetic neutron stars, as well
as the effects of their strong gravitational fields on the observable
properties. Using this model, we have analyzed the X-ray spectra of a
number of AXPs. We find that in all cases, the X-ray spectra are
described very well with this emission model. The spectroscopically
measured magnetic field strengths of these sources are in close
agreement with the values inferred from their spindown properties and
provide independent evidence for their magnetar nature. The analysis
of spectral data using this physical model also sheds light on the
long-term evolution of AXPs.

\end{abstract}


\maketitle


\section{Introduction}

Anomalous X-ray Pulsars (AXPs) are thought to be the observational
manifestations of a class of ultramagnetic ($B \gtrsim 10^{14}$~G)
neutron stars, also called magnetars (see Woods \& Thompson 2006 and
Kaspi 2006 for reviews on magnetars and AXPs). Among the numerous
spectral and timing properties of these isolated X-ray sources, two
stand out for our focus in these proceedings.  Their X-ray spectra are
soft but non-Planckian, traditionally described by empirical functions
such as a blackbody (kT $\sim$ 0.3$-$0.6 keV) plus a power law (with
photon index $\Gamma \sim$2.5$-$4) and, less frequently, by a sum of
two blackbody functions (see, e.g., Gotthelf \& Halpern 2005; Kaspi
2006).  The second property is their high spin-down rates, with
$\dot{P} \sim 10^{-11}$~s~s$^{-1}$.

A convincing, albeit indirect, argument for their strong magnetic
fields arises from these large spindown rates\footnote{The energetics
and the timescales of intense, super-Eddington, random bursts of
X-rays or soft gamma-rays seen in AXPs and the closely related Soft
Gamma-ray Repeaters that last a fraction of a second also suggest
independently the existence of very strong magnetic fields (Thompson
\& Duncan 1995).} (e.g., Kouveliotou et al.\ 1998).  Assuming the
neutron stars spin down due to magnetic braking of a dipole in vacuum,
their magnetic field strengths can be estimated by $B_{\rm dip} = 2.48
\times 10^{14}\; (P/6~{\rm s})^{1/2} (\dot{P}/10^{-11}~{\rm s}~{\rm
s}^{-1})^{1/2}$~G, for a neutron star moment-of-inertia $I =
10^{45}$~g~cm$^2$ and a neutron star radius of $R=10$~km. The dipole
fields associated with AXPs thus exceed $B \gtrsim 5 \times
10^{13}$~G. The dipole spindown formula makes numerous assumptions
when connecting period derivatives with a magnetic field strength,
such as a fiducial angle between the magnetic and rotation axes and
the absence of other torques on the neutron star (Spitkovsky 2006).
The dipole magnetic field inferred in this way has never been compared
with an independent, spectroscopic measurement for an isolated pulsar.

Recently, there has been significant theoretical work on the emission
from the atmospheres and magnetospheres of magnetars. As part of these
efforts, we have developed a physical model of emission from a
magnetar that takes into account processes in its atmosphere as well
as in its magnetosphere. The Surface Thermal Emission and
Magnetospheric Scattering (STEMS) model is based on the radiative
equilibrium atmosphere calculations presented in \"Ozel (2003) but
also includes the effects of magnetospheric scattering of the surface
radiation as discussed in Lyutikov \& Gavriil (2006) and G\"uver,
\"Ozel \& Lyutikov (2006). We also take into account the general
relativistic effects in the strong gravitational field of the neutron
star, making our models directly comparable to the wealth of spectral
and timing data on AXPs. Naturally, comparison with such data is the
ultimate test of any theoretical model.  At the same time, a model
that can describe consistently and in detail the spectra of AXPs can
be used to understand the physical properties of these sources and
their emission mechanisms.

In this proceedings paper, we present the results of applying the
Surface Magnetospheric Scattering and Surface Emission Model to the
soft X-ray data of four AXPs. In particular, we measure the magnetic
field strength of these sources spectroscopically and we investigate
the connection between the spectroscopically determined magnetic field
strengths with those inferred from dipole spindown.

\section{The Theoretical Model}

In highly magnetic, ionized neutron star atmospheres,
polarization-mode dependent transport of radiation that includes
absorption, emission, and scattering processes determines the
continuum spectrum (see, e.g., \"Ozel 2001, 2003). Furthermore, the
interaction of the photons with the protons in the plasma gives rise
to an absorption feature at the proton cyclotron energy $E_p = 6.3
\;(B/10^{15}{\rm G}) \; {\rm keV}$. This absorption feature is weakened
by the vacuum polarization resonance, which also leads to an enhanced
conversion between photons of different polarization modes as they
propagate through the atmosphere.

In the magnetospheres of magnetars, currents supporting the
ultrastrong magnetic fields can lead to enhanced charge densities
(Thompson, Lyutikov, \& Kulkarni 2002), which reprocess the surface
radiation through resonant cyclotron scattering (Lyutikov \& Gavriil
2006; G\"uver, \"Ozel, \& Lyutikov 2007). We calculate this effect
using the Green's function approach described in Lyutikov \& Gavriil
(2006) assuming that the magnetosphere is spherically symmetric and
the field strength follows a 1/$\rm{r}^{3}$ dependence.

In our spectral models, we include the relevant processes that take
place on the magnetar surface and its magnetosphere, which depend only
on four physical parameters. The first two parameters, the surface
magnetic field strength $B$ and temperature $T$, describe the
conditions found on the neutron star surface.  The third parameter
denotes the average energy of the charges $\beta = v_e/c$ in the
magnetosphere, while the last parameter is related to the density
$N_e$ of such charges and indicates the optical depth to resonant
scattering by $\tau = \sigma \int N_e dz$.  Here, $\sigma$ is the
cross-section for resonant cyclotron scattering.  We also assume a
fixed value for the gravitational acceleration on the neutron star
surface of $1.9\times10^{14}$~cm~s$^{-2}$, obtained for reasonable
values of the neutron star mass and radius.

We calculated model X-ray spectra (in the 0.05 - 9.8 keV range) by
varying model parameters in suitable ranges that are in line with the
physical processes we incorporated into the models: surface
temperature $T=0.1$ to $0.6$~keV, magnetic field $B=5\times10^{13}$ to
$3\times10^{15}$ G, electron velocity $\beta = 0.1$ to 0.5, and
optical depth in the magnetosphere $\tau =1$ to 10.  From the set of
calculated spectra, we created a table model which we use within the
X-ray spectral analysis package XSPEC (Arnaud 1996) to model the X-ray
spectra of AXPs.

Our models predict strong deviations from a Planckian spectrum, with a
hard excess that depends on the surface temperature as well as the
magnetic field strength, and weak absorption lines due to the proton
cyclotron resonance. Both the atmospheric processes and the
magnetospheric scattering play a role in forming these spectral
features and especially in reducing the equivalent widths of the
cyclotron lines.

\section{Analyses of AXP Spectra}

In this proceedings paper, we present the analysis of a total of four
XMM-Newton observations of four AXPs. For 4U~0142$+$61,
1RXS~J1708$-$4009, and XTE~J1810$-$197 we chose the longest available
X-ray observation carried out by XMM or Chandra observatories (i.e.,
the observation with the highest total counts). For 1E~1048.1$-$5937,
we used the longest observation of this source in quiescence. A longer
observation taken just after a burst from this source will be
presented elsewhere.

\begin{table}
\begin{tabular}{cccccccc}
\hline
    \tablehead{1}{c}{b}{Source}
  & \tablehead{1}{c}{b}{Satellite}
  & \tablehead{1}{c}{b}{Detector}
  & \tablehead{1}{c}{b}{Mode}
  & \tablehead{1}{c}{b}{Exposure\\Time (ks)}   
  & \tablehead{1}{c}{b}{Obs ID}   
  & \tablehead{1}{c}{b}{Obs Date}   
\\
\hline
4U~0142$+$61 & XMM-Newton & EPIC-PN & Fast Timing  & 21.1  & 0206670101 & Jul 25 2004 \\
1E~1048.1$-$5937 & XMM-Newton & EPIC-PN & Small Window & 32.44 & 0307410201 & Jun 16 2005 \\
1RXS~J1708$-$4009 & XMM-Newton & EPIC-PN & Small Window & 44.9  & 0148690101 & Aug 29 2003 \\
XTE~J1810$-$197 & XMM-Newton & EPIC-PN & Small Window & 42.2  & 0301270501 & Mar 18 2005 \\
\hline
\end{tabular}
\caption{Observations used for this study.}
\label{tab:obs}
\end{table}

In Table~\ref{tab:obs}, we present the list of the archival pointed
X-ray observations of each source analyzed in this study. All of these
observations were taken with the European Photon Imaging Camera (EPIC)
PN camera. The observations of 1E~1048.1$-$5937, 1RXS~J1708$-$4009,
and XTE~J1810$-$197 were taken in the Small Window Mode, while the
observation of 4U~0142$+$61 was taken in the Fast Timing Mode.


The spectral analysis was performed using the XSPEC 11.3.2.t (Arnaud
1996). We assumed a fiducial gravitational redshift correction of 0.2,
which corresponds to a neutron star with mass 1.4 $M_{\sun}$ and $R =
13.8$~km. We calculate the fluxes for the $0.5-8.0$~keV energy range
and quote errors for 90\% confidence level. 

\subsection{AXP 4U~0142$+$61}

4U~0142$+$61 is the brightest known AXP and has historically been very
stable. Rotating with a 8.69~s period (Israel et al. 1994), it spins
down at a rate of $\dot P \approx 0.196$~s~s$^{-1}$, yielding a $B_{\rm
dip} = 1.3 \times 10^{14}$~G using the dipole spindown formula.
Multiple X-ray observations of the source showed a long epoch of
nearly constant flux levels as well as a relatively hard X-ray
spectrum (Juett et al. 2002; Patel et al. 2003; G\"ohler, Wilms \&
Staubert 2005). Recently, the source exhibited SGR like bursts (Kaspi,
Dib \& Gavriil 2006; Dib et al. 2006; Gavriil et al. 2007) for the
first time.

4U 0142$+$61 has also been detected in hard X-rays with INTEGRAL
(Kuiper et al. 2006, den Hartog et al. 2007a). The hard X-ray spectral
component in the $20-230$~keV energy range is well described by a
power law model of index $0.79$ and the corresponding flux is $1.7
\times 10^{-10}$~erg~cm$^{-2}$s$^{-1}$ (den Hartog et al. 2007a), which
exceeds by a factor of $\sim$2 the unabsorbed 2-10 keV flux. If this
component extends without a break towards lower photon energies it
contributes significantly to the soft X-ray flux in the 7-10 keV
range.  Because of this, in our present analysis, we take into account
the effect of this component by using the fits to the hard X-ray
observations reported by den Hartog et al.\ (2007a), assuming that this
component extends to the soft X-rays without a break.

\subsection{1E~1048.1$-$5937}

Several properties distinguish 1E~1048.1$-$5937 from the other AXPs.
An ongoing RXTE monitoring campaign (Gavriil and Kaspi 2004) revealed
that it shows long-lived pulsed flux flares in addition to SGR-like
bursts. The spindown of the 6.452~s period pulsar is very unstable,
with period derivative values in the range $\dot{P} = 0.8546(50)-3.81
\times 10^{-11}$~s~s$^{-1}$ (Kaspi et al.\ 2001). This yields a large
range of dipole magnetic field strengths estimated from spindown.  As
a conservative range, we will adopt $B_{\rm dip} =2.4-4 \times
10^{14}$~G for this source.

1E~1048.1$-$5937 has been observed by the Chandra and XMM
observatories as part of ongoing campaigns to monitor the variability
of this AXP. The longest observation to date was taken by XMM on 16
June 2003, shortly after bursting activity. To focus on the quiescent
properties of this source, as with the other AXPs, we analyze here the
longest observation in quiescence, taken on 16 June 2005.

\subsection{1RXS~J1708$-$4009}

1RXS J1708$-$4009 is an 11.0~s AXP, initially thought to be a fairly
stable rotator (Israel et al. 1999). In the last several years, the
source experienced multiple glitches (e.g., Dib et al.\ 2007) that
interrupted stretches of steady spin-down. A period derivative of
$\dot{P} \approx 1.4-1.9 \times 10^{-11}$s~s$^{-1}$ yields a dipole
magnetic field strength of $B_{\rm dip}= 4.0-4.7 \times 10^{14}$~G.

As in the case of 4U~0142$+$61, 1RXS~J1708$-$4009 exhibits a hard,
pulsed hard X-ray tail extending to energies up to $\sim$150~keV
(Kuiper et al.\ 2006). Here we adopt the values given by den Hartog et
al. (2007b) with $\Gamma=1.17$ and the 20$-$250 keV flux 6.2$ \times
10^{-11}$ erg~cm$^{-2}$s$^{-1}$.

\subsection{XTE~J1810$-$197}

In the opposite extreme from 4U~0142$+$61, XTE~J1810$-$197 is the most
variable confirmed AXP observed to date. It was discovered (Ibrahim et
al.\ 2004) in 2003 when it suddenly brightened to more than 100 times
its quiescent value (Halpern \& Gotthelf 2005) during an outburst. The
source showed a steady decline of its X-ray flux thereafter, down to
unusually low quiescent flux levels that have been determined from
archival XTE and ROSAT data, accompanied by significant spectral
changes (Gotthelf \& Halpern 2006), earning it the title of the
transient AXP. The detection of characteristic X-ray bursts (Woods et
al.\ 2005), similar to those seen in other AXPs (Gavriil, Kaspi, \&
Woods 2002), further strengthen its classification as an AXP.

XTE~J1810$-$197 has a 5.54~s period, and an unsteady spindown
characterized by a $\dot P \approx 10^{-11}$~s~s$^{-1}$ period
derivative measured in the X-rays (Ibrahim et al.\ 2004; Gotthelf \&
Halpern 2005). The detection of radio emission from the source, first
ever for an AXP (Camilo et al.\ 2006), allowed for a more
closely-spaced monitoring of its period and resulted in the
measurement of a larger range of spindown rates (Camilo et al.\
2007). The range of $B_{\rm dip}$ corresponding to the observed
period derivatives are used in Figure~5.

In an earlier investigation, we reported on the physical evolution of
this source during its decline from outburst (G\"uver et al.\ 2007).
Here, we focus on the magnetic field strength of XTE J1810$-$197 
using the observation with the highest number of counts. 

\begin{figure}
   \includegraphics[height=.35\textheight,angle=-90]{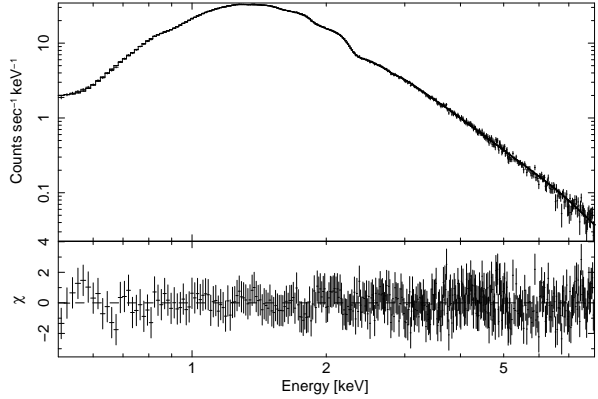}
   \caption{STEMS model fit to the X-ray spectrum of 4U~0142$+$61.}
\end{figure}

\begin{figure}
  \includegraphics[height=.35\textheight,angle=-90]{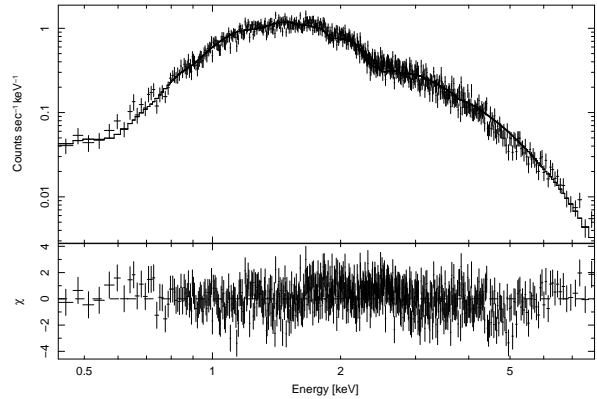}
  \caption{STEMS model fit to the X-ray spectrum of 1E~1048.1$-$5937.}
\end{figure}

\begin{figure}
  \includegraphics[height=.35\textheight,angle=-90]{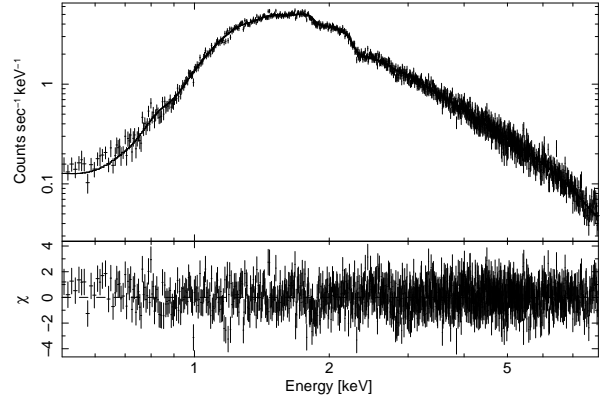}
  \caption{STEMS model fit to the X-ray spectrum of RXS~J1708$-$4009.}
\end{figure}

\begin{figure}
  \includegraphics[height=.35\textheight,angle=-90]{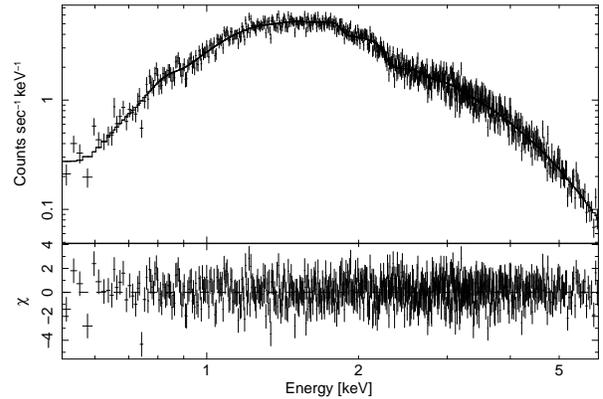}
  \caption{STEMS model fit to the X-ray spectrum of XTE~J1810$-$197.}
\end{figure}

\begin{table}
\begin{tabular}{cccccccc}
\hline
    \tablehead{1}{c}{b}{Source}
  & \tablehead{1}{c}{b}{Magnetic Field}
  & \tablehead{1}{c}{b}{Surface Temperature}
  & \tablehead{1}{c}{b}{$\tau$}
  & \tablehead{1}{c}{b}{$\beta$}
  & \tablehead{1}{c}{b}{$\chi^{2}_{\nu}$(d.o.f.)}   
\\
\hline
4U~0142$+$61 & 4.60$\pm$0.07 & 0.31$\pm 0.01$ & 3.54$\pm 0.14$ & 0.43 $\pm 0.01$ & 0.931 (462)\\
1E~1048.1$-$5937 & 2.26$\pm$0.05 & 0.37$\pm 0.01$ & 3.91$\pm 0.52$ & 0.22 $\pm 0.02$ & 0.952 (611) \\
1RXS~J1708$-$4009 & 3.96$\pm$0.17 & 0.35$\pm 0.01$ & 5.26$\pm 0.36$ & 0.48 $\pm 0.01$ & 1.050 (1187) \\
XTE~J1810$-$197 & 2.68$\pm$0.06 & 0.31$\pm 0.01$ & 2.36$\pm 0.36  $ & 0.25 $\pm 0.02$ & 1.07  (732) \\
\hline
\end{tabular}
\caption{Spectral Results of STEMS Model for 4 AXPs}
\label{tab:a}
\end{table}

\section{Discussion}

Figure~5 shows the magnetic field strengths obtained for the 4 AXPs by
fitting their X-ray spectra with the STEMS model against the dipole
field strengths inferred from the spindown of these sources according
to the dipole spindown formula. The error bars correspond to
$2-\sigma$ uncertainty in the values of the spectroscopic magnetic
field strength, while error bars on the dipole spindown field reflect
the range obtained from the variable period derivatives seen in some
sources.

\begin{figure}
  \includegraphics[height=.35\textheight]{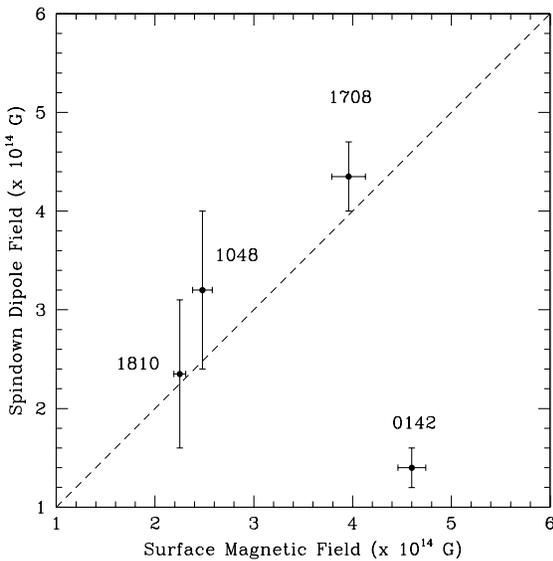}
  \caption{The comparison of the spectroscopically measured magnetic 
field strengths of five AXPs to the dipole fields inferred from the
spindown properties of these sources. The error bars in $B_{\rm dip}$ 
represent the range of measured spindown rates for each source, while 
the error bars in the spectroscopic magnetic field strength represent
$2-\sigma$ statistical uncertainties.}
\end{figure}

In the cases of 1E~1048.1$-$5937, 1RXS~J1708$-$4009, and
XTE~J1810$-$197, we find a very good agreement, at an unexpected
level, between the spectroscopically measured magnetic field strength
and that obtained from their spindown. In all three cases, the two
values are consistent within the formal and expected systematic
uncertainties. For the case of 4U~0142$+$61, the spectroscopically
measured surface magnetic field is a factor of 3 larger than the
spindown field. This may be due to the simplified assumptions in
either of the two measurements of the magnetic field.  Alternatively,
this might be an indication of multipole magnetic field components on
the neutron star surface which contribute negligibly to the spindown
torques.

The good agreement between the theoretical models and the spectral
data provide us with a new tool with which to understand the physical
conditions of magnetar surfaces and magnetospheres. At the same time,
our spectroscopic measurements of the magnetic field strengths offer
independent confirmation for the magnetar nature of AXPs.


\begin{theacknowledgments}
F.O. acknowledges support from NSF grant AST-0708640.  We thank the
McGill Pulsar group, and in particular, C. Tam, for maintaining the
online AXP/SGR catalog
\url{http://www.physics.mcgill.ca/~pulsar/magnetar/main.html} that was
helpful in preparing this publication.

\end{theacknowledgments}






\end{document}